\documentclass[dvips,a4paper,11pt]{article}

\usepackage{array}
\usepackage{dcolumn}
\usepackage[dvips]{graphicx}
\usepackage{amsopn}
\usepackage{amssymb,amsfonts,amsbsy}
\usepackage{psfrag}
\usepackage{latexsym}

\usepackage{array}
\usepackage{dcolumn}

\usepackage{epsfig}
\usepackage{enumerate,cite,color}
\usepackage{psfrag}
\usepackage{graphics}
\usepackage{amsmath,amsfonts,amsthm}

\usepackage{sectsty}

\usepackage{url}

\def\Rs{\mathbb{R}}
\def\Ns{\mathbb{N}}

\theoremstyle{definition}

\title{Orthogonal systems of Zernike type in polygons and polygonal facets}
\author{\\Chelo Ferreira$^1$, Jos\'{e} L. L\'{o}pez$^2$, Rafael Navarro$^3$  and Ester P\'{e}rez Sinus\'{\i}a$^1$}
\date{}

\begin{document}
\normalsize \maketitle

{\small
\begin{center}
$^1$ \textsf{\textit{Dpto. de Matem\'{a}tica Aplicada, IUMA, Universidad de Zaragoza}}\\
\textsf{\textit{e-mail: cferrei@unizar.es,
ester.perez@unizar.es}}\\
$^2$\textsf{\textit{Dpto. de Ingenier\'{\i}a Matem\'{a}tica e Inform\'{a}tica, Universidad P\'{u}blica de Navarra and BIFI, Zaragoza}}\\
\textsf{\textit{ e-mail: jl.lopez@unavarra.es}}\\
$^3$\textsf{\textit{ICMA, Consejo Superior de Investigaciones CientÆficas \& Universidad de Zaragoza, Zaragoza}}\\
\textsf{\textit{ e-mail: rafaelnb@unizar.es}}
\end{center}}

\begin{abstract}
Zernike polynomials are commonly used to represent the wavefront phase on circular optical apertures, since they form a complete and orthonormal basis on the unit disk. In [Diaz et all, 2014] we introduced a new Zernike basis for elliptic and annular optical apertures based on an appropriate diffeomorphism between the unit disk and the ellipse and the annulus. Here, we present a generalization of this Zernike basis for a variety of important optical apertures, paying special attention to polygons and the polygonal facets present in segmented mirror telescopes. On the contrary to ad hoc solutions, most of them based on the Gram-Smith orthonormalization method, here we consider a piece-wise diffeomorphism that transforms the unit disk into the polygon under consideration. We use this mapping to define a Zernike-like orthonormal system over the polygon. We also consider ensembles of polygonal facets that are essential in the design of segmented mirror telescopes. This generalization, based on in-plane warping of the basis functions, provides a unique solution, and what is more important, it guarantees a reasonable level of invariance of the mathematical properties and the physical meaning of the initial basis functions. Both, the general form and the explicit expressions for a typical example of telescope optical aperture are provided.
 \\\\
\noindent \textsf{2010 AMS \textit{Mathematics Subject
Classification:} 41A10 (Approximation by polynomials); 41A63 (Multidimensional problems); 78M99 (None of the above, but in this section: Optics, electromagnetic theory); 85-08 (Computational methods: Astronomy and astrophysics).
} \\\\
\noindent  \textsf{Keywords \& Phrases:} Zernike polynomials; orthonormal systems; polygonal facets; segmented mirror telescopes.
\\\\
\end{abstract}

\section{Introduction}

The problem of finding complete orthonormal systems to represent functions defined on finite supports with a given geometry appears in many areas of Physics and Engineering. In particular, Zernike circle polynomials \cite{zernike} are widely used to represent optical path differences (phase differences or wave aberrations) in wavefronts, or even the sag of optical surfaces (such as the human cornea \cite{corneauno}, \cite{corneados}) as they are well adapted to the circular shape (disks or slightly deformed disks) of a majority of conventional optical systems. There is an infinite number of possible systems, but Zernike polynomials (ZPs) (or lineal combinations of them \cite{luk}) show important advantages and interesting properties. Among these properties, ZPs permit to establish a link with the traditional Seidel theory of aberrations \cite{born}, which is based on a third order Taylor series expansion, and with further extensions of the Seidel theory to 5th order, etc. The theoretical and practical advantages of orthonormal polynomials, make that ZPs became the standard way to describe the phase of wavefronts \cite{nij} (or the wave aberration or optical path differences) in many fields ranging from atmospheric optics \cite{noll}, optical design and testing \cite{mala} or visual optics (the ANSI Z80.28 standard for reporting aberrations in the human eye is based on ZPs).

The circle is the most common optical aperture, but other more complicated geometries have gained interest in recent years after the development of large telescopes that require more sophisticated optical designs such as the annular pupils in large telescopes \cite{resta}. In particular, segmented mirror telescopes (SMTs) are commonly used nowadays by NASA, ESA and other astronomical and astrophysical organizations all over the world to collect information from the outer space in the form of electromagnetic radiation. An SMT is an array of smaller mirrors designed to act as segments (or facets) of a single large mirror. The facets can have diverse shapes, ranging from planar squares to curved asymmetric polygons \cite{patrick}. They are used as objectives for large reflecting telescopes. To function, all the mirror segments have to be polished to a precise shape and actively aligned by a computer controlled active optics system using actuators built into the mirror support cell.


An important example of SMT is The European Extremely Large Telescope \cite{european}. Because current monolithic mirrors cannot be constructed larger than about eight meters in diameter, the use of segmented mirrors is a key component of current large-aperture telescopes. This is because of the technological limit of a primary mirror made of a single rigid piece of glass \cite{nick}. Using a monolithic mirror much larger than 5 meters is prohibitively expensive due to the cost of both the mirror, and the massive structure needed to support it. A mirror beyond that size would also sag slightly under its own weight as the telescope was rotated to different positions changing the precision shape of the surface \cite{monica}. Facets are also easier to fabricate, transport, install, and maintain over very large monolithic mirrors.

Other examples of SMTs are: (i) The Keck Telescope in Hawaii, the largest optical and infrared telescope in the world, that consists of 36 hexagonal segments, each 0.9 meters on a side (http://www.ps.uci.edu/physics/news/chanan.html). (ii)
The {\it Gran Telescopio Canarias} (GTC), whose useful wavelength range extends from 365nm to 25 $\mu$m, its primary mirror consists of 36 hexagonal facets of Zerodur coated with aluminium; each segment measures 1.9m from vertex to vertex and has a side length of 0.9m and a weight of 470 kg. They cover a total surface of 75.7 square metres and have gaps of 3mm between them (\url{http://www.gtc.iac.es/observing/GTCoptics.php$\#$M1}).

\begin{figure}[h!]
\begin{center}
  \includegraphics[width=4in]{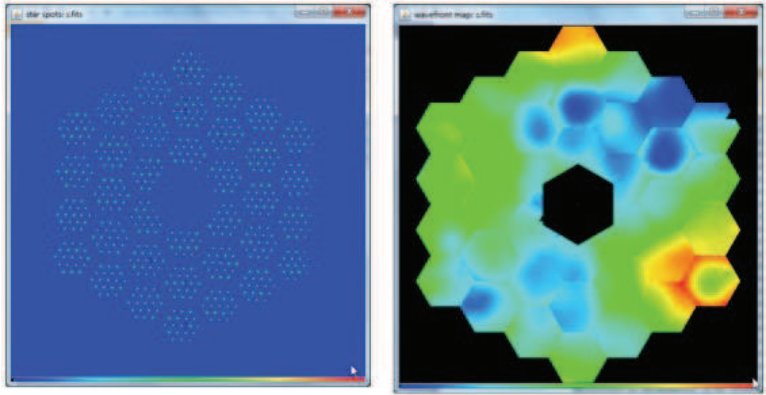}\\
  \caption{Example image of a wavefront test (left) and example of wavefront solution (right). Pictures taken from the GTC theoretical base data: http://www.gtc.iac.es/observing/GTCoptics.php$\#$M1.}
  \end{center}
\end{figure}

The importance of these non-circular geometries motivated the development of a series of ad hoc solutions, most of them based on the Gram-Smith (G-S) method to obtain orthonormal basis on different types of apertures \cite{swan}, \cite{up} such as ellipses \cite{diaz}, rectangles \cite{majandos}, annuli \cite{majantres}, circular sectors \cite{diazdos}, etc. The main drawback of the G-S method is that it does not provide of a unique solution that hinder the physical interpretation of the associated expansion coefficients (especially for the higher orders due to a cumulative effect associated to the G-S method).

In our previous paper \cite{nuestro}, we proposed a mapping that transforms the unit circle into the desired geometry, paying special attention to elliptic and annular aperture geometries. Using this mapping, we obtained a new Zernike basis for elliptic and annular optical apertures with important mathematical and physical properties.
In this paper we go a step further giving a rigorous formulation under a unique criterion and providing a unique general solution for most of the usual optical apertures. As it is explained in \cite{nuestro}, our approach is based in finding an appropriate mapping that transforms the unit circle into the desired aperture geometry. That is finding piece-wise diffeomorphism that transforms the unit circle into the set within the plane that represents the optical aperture. This mapping means warping the input basis functions so that they fit into the new aperture geometry. On the contrary that ad hoc solutions, that warping permits not only unicity, but also a high level of invariance of the mathematical properties and physical meaning of the basis functions (tilt, defocus, astigmatism, coma, etc.) and hence a natural generalization of the aberration theory.

Although this technique may be applied to different optical systems defined by different geometries (ellipses, circular sectors, annuli,...), in this paper we focus our attention in the geometries required for the SMTs: polygons and ensembles of polygonal facets. Apart from SMTs, a different optical context where this kind of geometry is of interest can be found in the high resolution compound superposition eyes of some diurnal insects that suggest high resolution optics \cite{navarro}.

In the following section we report the definition of the Zernike orthonormal system and revise, within a rigorous formulation, the general method introduced in \cite{nuestro}, that is valid for any domain, piece-wise diffeomorphic to the unit disk. In Section 3 we focus our attention on polygonal domains and give a detailed description of the corresponding Zernike-like system. In Section 4 we consider an ensemble of hexagons typically used in the primary phase of SMTs and give an orthonormal system composed by a combination of the Zernike-like systems of the facets. In Section 5 we give an example of wave front description in a polygon and in an ensemble of polygonal facets using the Zernike-like basis introduced in Sections 3 and 4. Some final remarks are given in Section 6.

\section{Zernike-like systems in compact sets of the plane}

\subsection{Zernike polynomials in the unit disk}

Through out this paper, $\vec{u}:=(u,v)$ represent cartesian coordinates in the unit Disk $D$ and $\rho$, $\phi$ the corresponding polar coordinates: $(u,v)=\sigma(\rho,\phi):=(\rho\cos\phi,\rho\sin\phi)$, $0\le\rho^2=u^2+v^2\le 1$ and $0\le\phi=\arctan(y/x)<2\pi$. Also, we will use the notation $Z_n^m(u,v)$ for the Zernike polynomials in cartesian coordinates and the notation $\bar Z_n^m(\rho,\phi)$ for the Zernike polynomials in polar coordinates defined in the unit disk $D$: $\bar Z_n^m:=Z_n^m\circ\sigma$. The standard definition of the Zernike polynomials in polar coordinates is the following \cite{nij}:
\begin{equation}\label{defpolares}
\bar Z_n^m(\rho,\phi):=
\left\lbrace
\begin{array}{ll}
\displaystyle N_n^m R_n^{|m|}(\rho)\cos(m\phi), & \hskip 5mm m\geq 0,\cr
\displaystyle -N_n^m R_n^{|m|}(\rho)\sin(m\phi), & \hskip 5mm m< 0.\cr
\end{array}
\right.
\end{equation}

In this formula, the polynomials $R_n^{|m|}(\rho)$ and the normalization factor $N_n^m$ are:
\begin{equation}\label{radial}
R_n^{|m|}(\rho):=\sum_{k=0}^{{(n-|m|)/2}}{(-1)^k (n-k)!\over k!\left({n+|m|\over 2}-k\right)!\left({n-|m|\over 2}-k\right)!}\rho^{n-2k}, \hskip 2cm
N_n^m:=\sqrt{{2(n+1)\over 1+\delta_{m,0}}}
\end{equation}
where $\delta_{m,0}$ is the Kronecker delta function: $\delta_{0,0}=1$ and $\delta_{m,0}=0$ for $m\neq 0$.
The index $n$ indicates the degree of the radial polynomial $R_n^{|m|}(\rho)$ and the index $m$ the azimutal frequency of the azimutal component. The index $n$ is any positive integer or zero: $n=0,1,2,...$. For a given $n$, the index $m$ takes the values $m=-n,-n+2, -n+4$, $\ldots,n$. More details, as the expression of the Zernike polynomials in cartesian coordinates may be found in \cite{nuestro} for example. For later convenience, we summarize here the orthogonality and completeness properties of the Zernike polynomials:

The family $\left\lbrace Z_n^m\right\rbrace_{n=0,1,2,...}^{m=-n,-n+2,...,n}$ is an orthonormal system in $L^2(D)$. Here, and in the remaining of the paper, $L^2(D)$ represents the set of square integrable functions defined on the unit disk $D$ with respect to the normalized Lebesgue measure $d\mu:=dudv/\pi$. Equivalently, the family $\left\lbrace \bar Z_n^m\right\rbrace_{n=0,1,2,...}^{m=-n,-n+2,...,n}$ is an orthonormal system in $L^2(D)$ with respect to the measure $d\bar\mu:=\rho d\rho d\phi/\pi$ [\cite{nij}, chap. 2]:
$$
(Z_n^m,Z_{n'}^{m'})_\mu:=\int\int_D Z_n^m\cdot Z_{n'}^{m'}d\mu=(\bar Z_n^m,\bar Z_{n'}^{m'})_{\bar\mu}:=\int\int_D \bar Z_n^m\cdot\bar Z_{n'}^{m'}d\bar\mu=\delta_{n,n'}\delta_{m,m'}.
$$

Moreover, the Zernike set is a complete orthomormal system of $L^2(D)$. That is, any square integrable function $f:D\to \mathbb{C}$ can be written in the form [\cite{nij}, chap. 2]:
\begin{equation}\label{suma}
f=\sum_{n=0}^\infty\sum_{m=-n}^n c_n^mZ_n^m, \hskip 2cm \bar f=\sum_{n=0}^\infty\sum_{m=-n}^n c_n^m\bar Z_n^m,
\end{equation}
where $\bar f:=f\circ\sigma$ is the function $f$ in polar coordinates. In the above equalities:
\begin{equation}\label{cnm}
c_n^m:=(Z_n^m,f)_\mu=\int\int_D Z_n^m\cdot f\,d\mu,\hskip 1cm {\rm or}\hskip 1cm c_n^m:=(\bar Z_n^m,\bar f)_{\bar\mu} =\int\int_D \bar Z_n^m\cdot\bar f\,d\bar\mu,
\end{equation}
The equality \eqref{suma} must be understood in the $L^2$ sense, that is, the convergence of the series in the right hand side is understood in the $L^2(D)$ norm $\vert\vert f\vert\vert^2:=(f,f)_\mu=(\bar f,\bar f)_{\bar\mu}$:
$$
\lim_{N\to\infty}\left\vert\left\vert f-\sum_{n=0}^N\sum_{m=-n}^n c_n^mZ_n^m\right\vert\right\vert=\lim_{N\to\infty}\left\vert\left\vert \bar f-\sum_{n=0}^N\sum_{m=-n}^n  c_n^m\bar Z_n^m\right\vert\right\vert=0.
$$

\subsection{Zernike-like systems in other sets of the plane}

Any planar optical aperture may be mathematically described as a set $M$ in the plane piece-wise diffeomorphic to the unit disk $D$. Then, we consider all the subsets $M\subset\Rs^2$ obtained from $D$ by means of a mapping $\varphi:D\to M$ that is a diffeomorphism a.e. in $D$ (see Figure 3):
\begin{equation}\label{dife}
\begin{array}{ll}
\varphi: &D\longrightarrow M\subset\Rs^2,\cr
&(u,v)\to(x,y)=\varphi(u,v). \cr
\end{array}
\end{equation}

\begin{figure}[h!]
\begin{center}
  \includegraphics[width=2.7in]{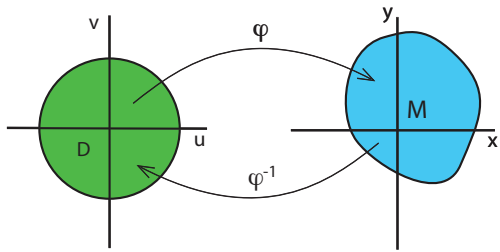}\\
  \caption{The unit disk $D$ and the set $M$ are deformed each in other by the respective mappings $\varphi$ and $\varphi^{-1}$.}
  \end{center}
\end{figure}

In the remaining of the paper we use the notation $\vec{x}:=(x,y)$ for the cartesian coordinates in $M$ and $r^2=x^2+y^2$ and $\theta=\arctan(y/x)$ for the corresponding polar coordinates in $M$, with $0\le\theta<2\pi$. Then, for $(x,y)\in M$ we write
$\vec{x}:=(x,y):=\varphi(u,v)=(x(u,v),y(u,v))$. And reciprocally, $\vec{u}:=(u,v)=\varphi^{-1}(x,y)=(u(x,y),v(x,y))$. We denote by $J(x,y)$ the jacobian of $\varphi^{-1}$ that is defined a.e in $M$:
$$
J(x,y):=\left({\partial\vec{u}\over\partial\vec{x}}\right),
$$
and $d\mu_J:=\vert J(x,y)\vert dxdy/\pi$ the measure induced by $\varphi$ in $M$. In the remaining of the paper we assume that $J(x,y)$ is continuous a. e. in $M$. Then, it is clear that:
$$
\delta_{n,n'}\delta_{m,m'}=\int\int_D Z_n^mZ_{n'}^{m'}d\mu=\int\int_M (Z_n^m\circ\varphi^{-1})\cdot(Z_{n'}^{m'}\circ\varphi^{-1})\,d\mu_J.
$$
In other words, the set of functions
$$
K_n^m:=Z_n^m\circ\varphi^{-1}; \hskip 2cm n=0,1,2,...; \hskip 1cm m=-n,-n+2, -n+4, \ldots,n,
$$
defined in $M$, is an orthonormal system of $L^2(M)$ of complex square integrable functions defined over $M$ with respect to the measure $d\mu_J$.

It is straightforward to see that the family $\lbrace K_n^m\rbrace_{n=0,1,2,...}^{m=-n,-n+2,...n}$ is complete in $L^2(M)$ with respect to the measure $d\mu_{J}$. For any function $F\in L^2(M)$, we define the function $f\in L^2(D)$ induced by the mapping $\varphi$ in the form $f:=F\circ\varphi$.
The Zernike system is a complete orthonormal system in $L^2(D)$ and then
$$
F\circ\varphi=\sum_{n=0}^\infty\sum_{m=-n}^n c_n^m Z_n^m
$$
a.e., with $c_n^m$ given in \eqref{cnm}. Using that $Z_n^m=K_n^m\circ\varphi$ we find that
\begin{equation}\label{exq}
F=\sum_{n=0}^\infty\sum_{m=-n}^n c_n^mK_n^m
\end{equation}
a.e., with
\begin{equation}\label{cj}
c_n^m=(K_n^m,F)_{d\mu{J}}:=\int\int_M K_n^m\cdot  F\,d\mu_{J}.
\end{equation}
\section{Zernike-like systems in polygons}

In this section we particularize the above general formulation to polygonal domains, that is, the set $M$ is a polygon of $p\ge 3$ sides and radius $R_0$ centered at the origin.
We define $\alpha:=\pi/p$ and assume that one of the radius of the polygon is located at an angle $\alpha$ from the positive X axis, as it is indicated in right picture of Figure 4 for the particular case $p=8$.

\begin{figure}[h!]
\begin{center}
\includegraphics[width=4in]{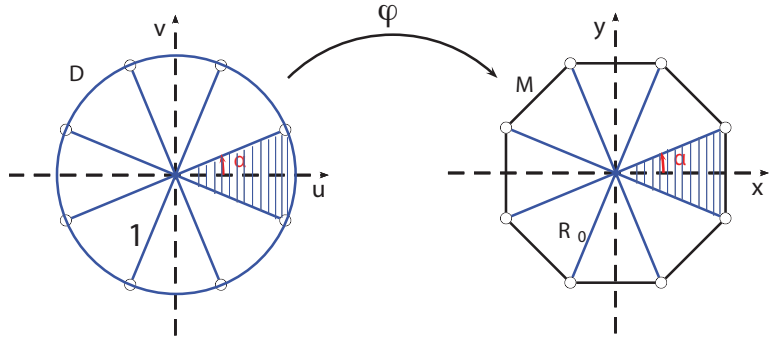}\\
  \caption{Right picture: octagon of $p=8$ sides centered at the origin. One of the sides is orthogonal to the X axis and is divided in two equal segments by the positive X axis. Left picture: division of the unit disk into $p$ identical sectors.}
 \end{center}
\end{figure}

The key point here is just to write down a convenient mapping $\varphi$ from the disk $D$ to the polygon $M$. To this end, we first find the mapping $\varphi_0$ from the sector of the disk $S_\alpha:=\lbrace(\rho\cos\phi,\rho\sin\phi)$; $-\alpha\le\phi<\alpha$, $0\le\rho\le 1\rbrace\subset D$ into the triangular portion of the polygon $M$ located over the positive X axis; the triangle of vertexes $(0,0)$ and $(R_0\cos\alpha, \pm R_0\sin\alpha)$. ($\varphi_0$ maps the shaded sector of the disk D into the shaded triangle of the octogonon $M$ in Figure 4.) This mapping $\varphi_0$ is given by
$$
(x,y)=(r\cos\theta,r\sin\theta)=\varphi_0(\rho,\phi)=R_0\rho\cos\alpha\left(1,{\sin\phi\over\cos\phi}\right),
\hskip 1cm 0\le\rho\le 1, \hskip 1cm -\alpha\le\phi<\alpha,
$$
where, for convenience, we have used polar coordinates $(\rho,\phi)$ in the disk. Observe that $\theta=\phi$ and $r=R_0\rho\cos\alpha/\cos\phi$.

Now, in order to find the complete mapping $\varphi:D\to M$, we just need to consider a division of the disk $D$ into $p$ sectors of amplitude $2\alpha$ and repeat the mapping $\varphi_0$ from everyone of these $p$ sectors to everyone of the $p$ triangles that compose the polygon $M$ (see Figure 4). To this end we define the function
$$
U_\alpha(\phi):=\phi-\left\lfloor{\phi+\alpha\over 2\alpha}\right\rfloor 2\alpha, \hskip 2cm 0\le\phi<2\pi.
$$
This function is piece-wise linear with slope 1, discontinuous at $\theta=(2m+1)\alpha$, for integer $m$ and bounded by $\pm\alpha$. We also define the variable radius
$$
R_\alpha(\phi):={R_0\cos\alpha\over\cos(U_\alpha(\phi))}.
$$
We find that $R_0\cos\alpha\le R_\alpha(\phi)\le R_0$. The function $R_\alpha(\phi)$ is an increasing function for $\phi\in(2m\alpha,(2m+1)\alpha)$ and decreasing for $\phi\in((2m+1)\alpha,2(m+1)\alpha)$, $m=0,1,2,...$.

With these preparations, we can finally define the mapping $\varphi$ that transforms the unit disk $D$ into the polygon $M$:
$$
(x,y)=\varphi(u,v)=(uR_\alpha(\arctan(v/u)),vR_\alpha(\arctan(v/u))),\hskip 2cm u^2+v^2\le 1.
$$
Or, in polar coordinates:
$$
r(\rho,\phi)=\rho R_\alpha(\phi),\hskip 2cm \theta(\rho,\phi)=\phi, \hskip 2cm 0\le\rho\le 1,\hskip 3mm 0\le\phi<2\pi.
$$
The jacobian of the inverse transformation
$$
(u,v)=\varphi^{-1}(x,y)=\left({x\over R_\alpha(\arctan(y/x))},{y\over R_\alpha(\arctan(y/x)}\right),
$$
is, in polar coordinates, $J_\alpha(\theta)=[R_\alpha(\theta)]^{-2}$. From Section 2 we find that the family $\lbrace K_n^m\rbrace_{n=0,1,2,...}^{m=-n,-n+2,...,n}$,
\begin{equation}\label{kpolycar}
K_n^m(x,y)=Z_n^m\left({x\over R_\alpha(\arctan(y/x))},{y\over R_\alpha(\arctan(y/x))}\right),
\end{equation}
or, in polar coordinates,
\begin{equation}\label{kpolypol}
\bar K_n^m(r,\theta)=\bar Z_n^m\left({r\over R_\alpha(\theta)},\theta\right),
\end{equation}
is a complete orthonormal system in $L^2(M)$ with respect to the measure (in cartesian and polar coordinates):
\begin{equation}\label{medida}
d\mu_J={dxdy\over\pi R^2_\alpha(\arctan(y/x))}, \hskip 3cm d\bar\mu_J={rdrd\theta\over\pi R^2_\alpha(\theta)}.
\end{equation}

More precisely:
\begin{equation}\label{barrak}
\bar K_n^m(r,\theta):=
\left\lbrace
\begin{array}{ll}
N_n^m R_n^{|m|}\left({r\over R_\alpha(\theta)}\right)\cos(m\theta), &\hskip 5mm m\geq 0,\cr
-N_n^m R_n^{|m|}\left({r\over R_\alpha(\theta)}\right)\sin(m\theta), &\hskip 5mm m< 0,\cr
\end{array}
\right.
\end{equation}
where the polynomials $R_n^{|m|}$ are given in \eqref{radial}. Also, the family $\lbrace H_n^m\rbrace_{n=0,1,2,...}^{m=-n,-n+2,...,n}$, with
\begin{equation}\label{qpolycar}
H_n^m(x,y)={1\over R_\alpha(\arctan(y/x))}Z_n^m\left({x\over R_\alpha(\arctan(y/x))},{y\over R_\alpha(\arctan(y/x))}\right),
\end{equation}
or, in polar coordinates,
\begin{equation}\label{qpolypol}
\bar H_n^m(r,\theta)={1\over R_\alpha(\theta)}\bar Z_n^m\left({r\over R_\alpha(\theta)},\theta\right),
\end{equation}
is a complete orthonormal system in $L^2(M)$ with respect to the measure (in cartesian and polar coordinates)
$$
d\mu={dxdy\over\pi}, \hskip 3cm d\bar\mu={rdrd\theta\over\pi}.
$$

Then, any function $F\in L^2(M)$ can be written, a.e., in the form
$$
F=\sum_{n=0}^\infty\sum_{m=-n}^n c_n^m K_n^m,
$$
with
$$
c_n^m={1\over \pi}\int\int_M K_n^m(x,y)F(x,y){dxdy\over R^2_\alpha(\arctan(y/x))}.
$$
Also, any function $F\in L^2(M)$ can be written, a.e., in the form
$$
F(x,y)=\sum_{n=0}^\infty\sum_{m=-n}^n c_n^m H_n^m,
$$
with
$$
c_n^m={1\over \pi}\int\int_M H_n^m(x,y)F(x,y)dxdy.
$$

Several representative examples, corresponding to various Zernike wavefront aberrations:  tilt, defocus, astigmatism, coma, trefoil and spherical aberration are represented in Figure 5 for the particular case of the octagon, $\alpha = \pi/8$ and $R_0 = 1$  (Only positive values of m are shown since $m < 0$ are rotated versions of the same aberration modes).

\begin{figure}[h!]
\begin{center}
\includegraphics[width=1.5in]{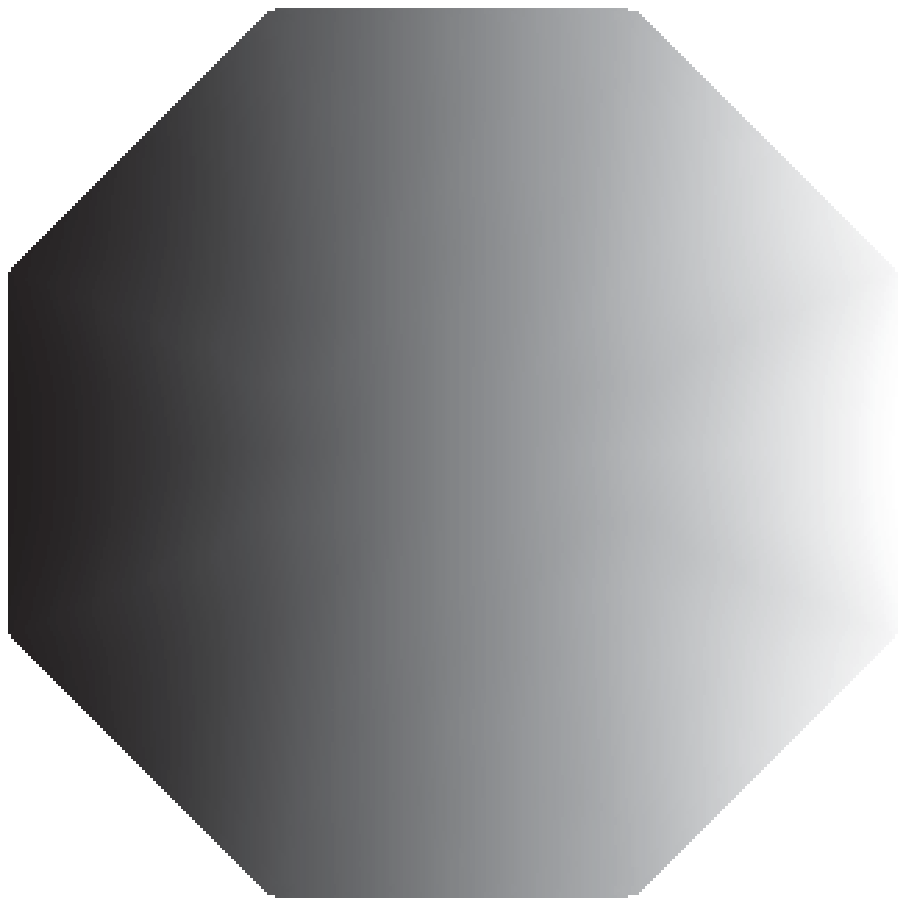}\quad\includegraphics[width=1.5in]{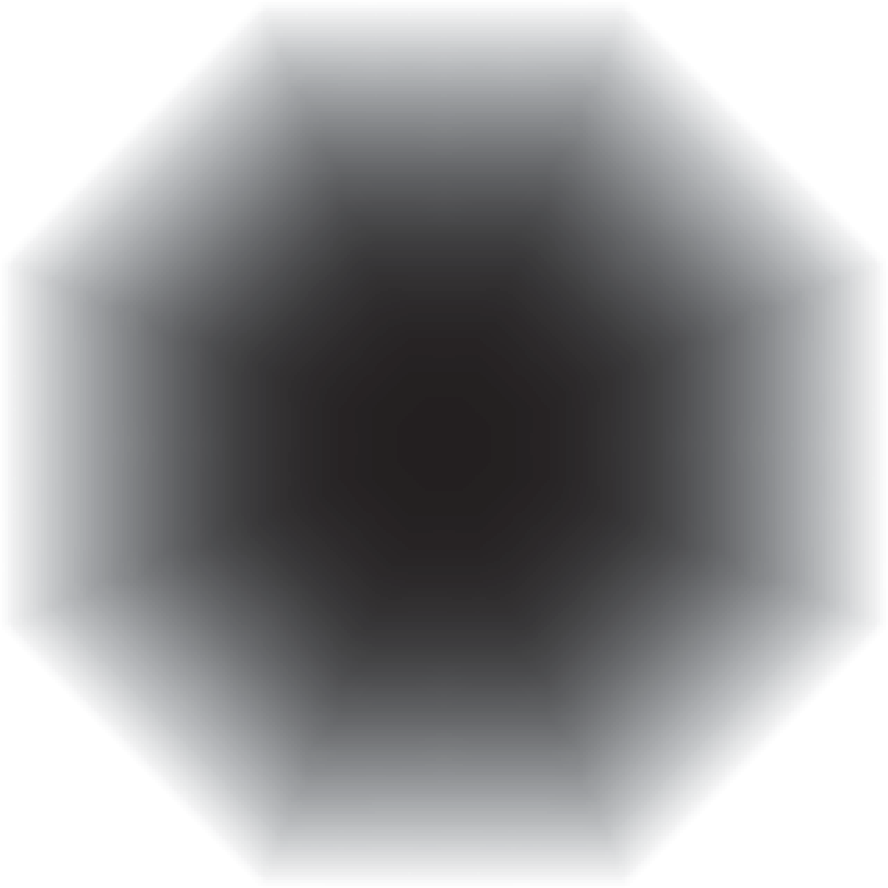}\quad\includegraphics[width=1.5in]{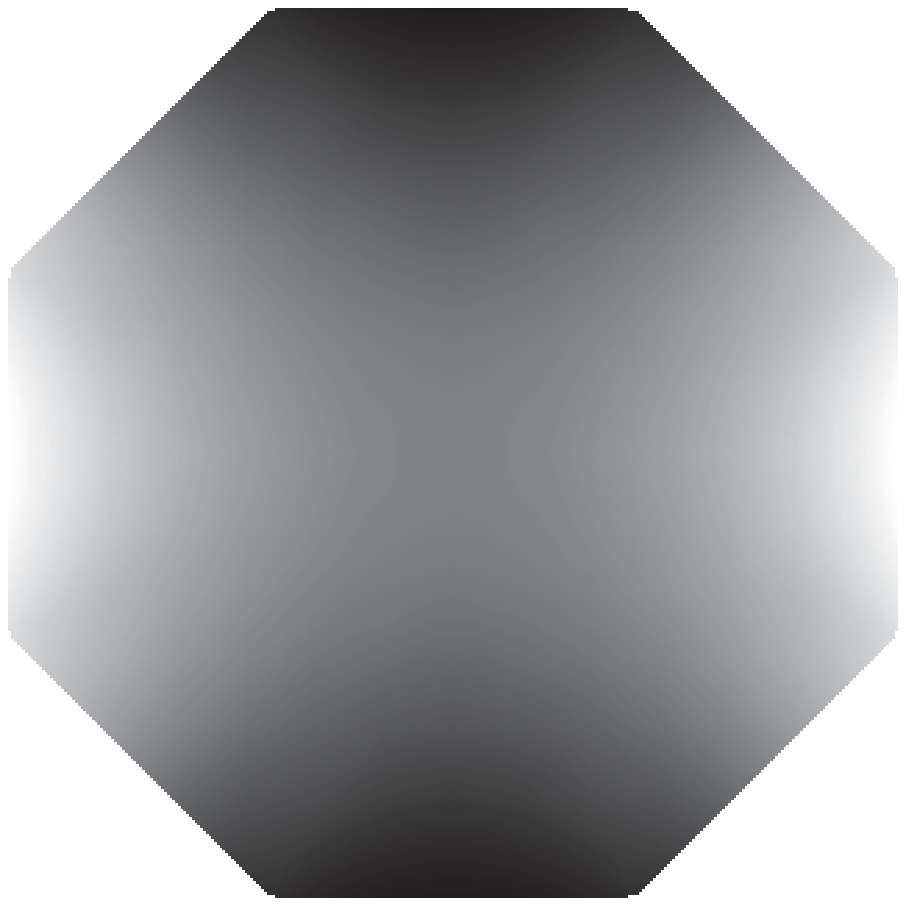}\\
\includegraphics[width=1.5in]{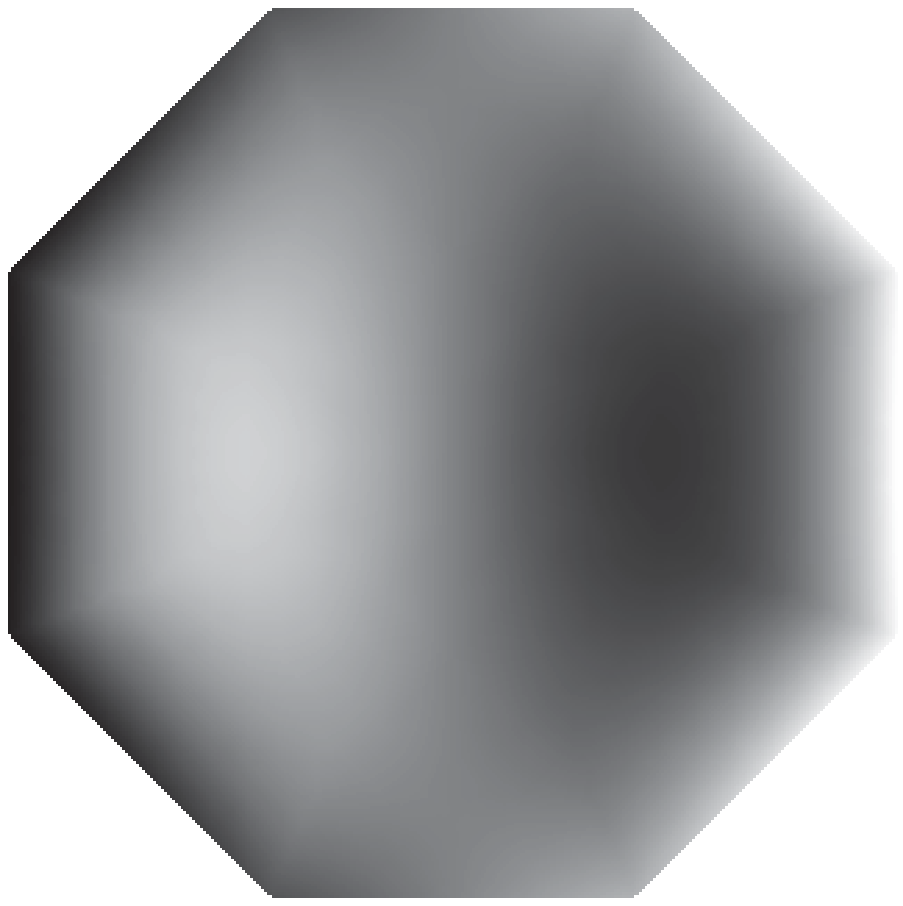}\quad\includegraphics[width=1.5in]{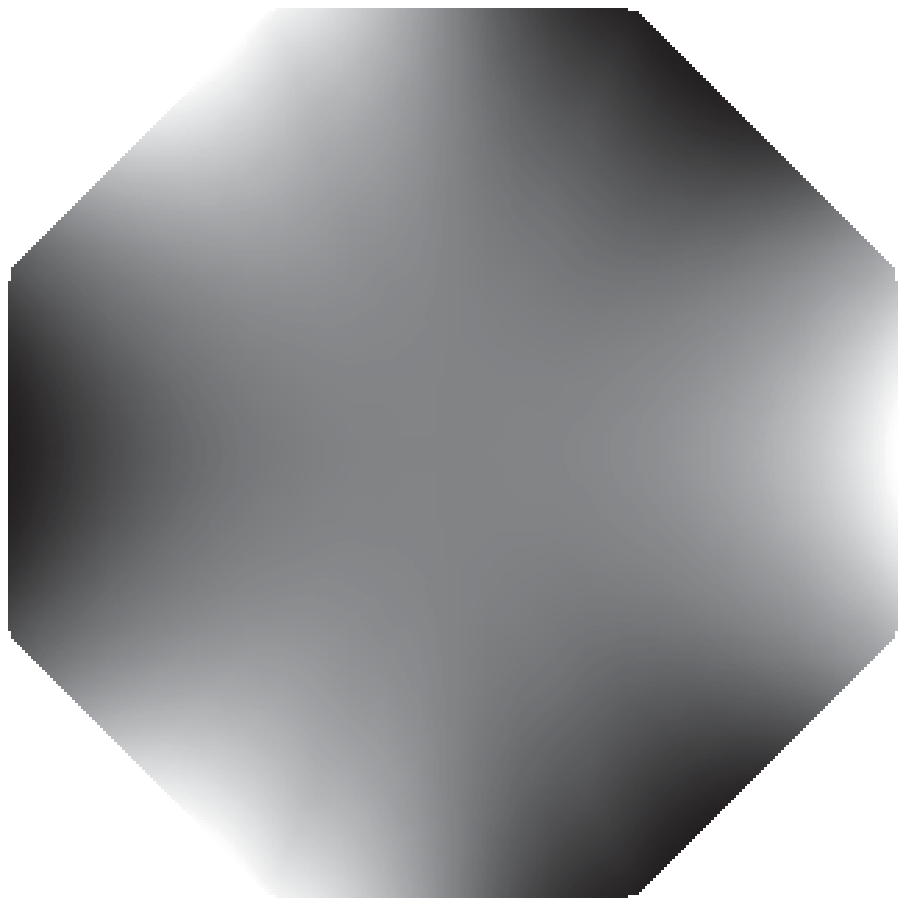}\quad\includegraphics[width=1.5in]{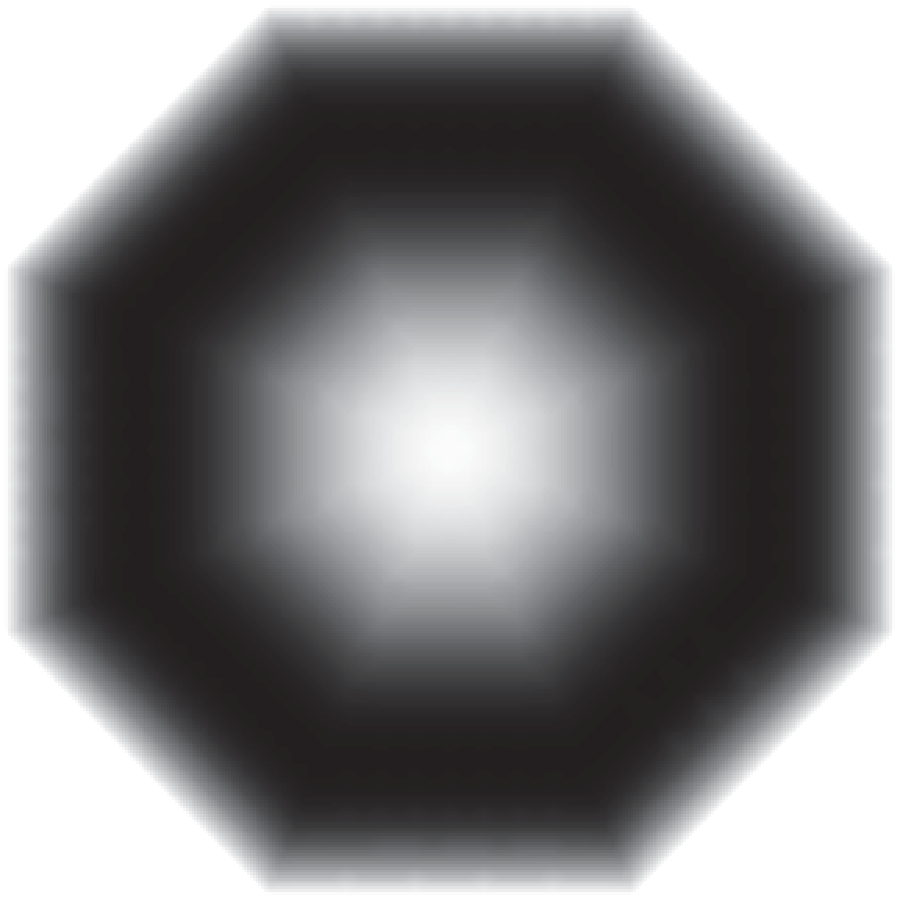}\\
  \caption{Level sets for the first few polygonal Zernike functions $K_n^m(x,y)$ in an octagon, that determine the first classical aberrations.}
  \end{center}
\end{figure}

\section{Zernike-like systems in an ensemble of hexagonal facets}

In this section we consider a set $A$ composed by the union of $N$ equal polygons of $p$ sides. A typical example frequently used in the primary phase of SMTs telescopes consist of 18 hexagons $M_k$, $k=1,2,...,18$ disposed as in Figure 6.

\begin{figure}[h!]
\begin{center}
\includegraphics[width=3in]{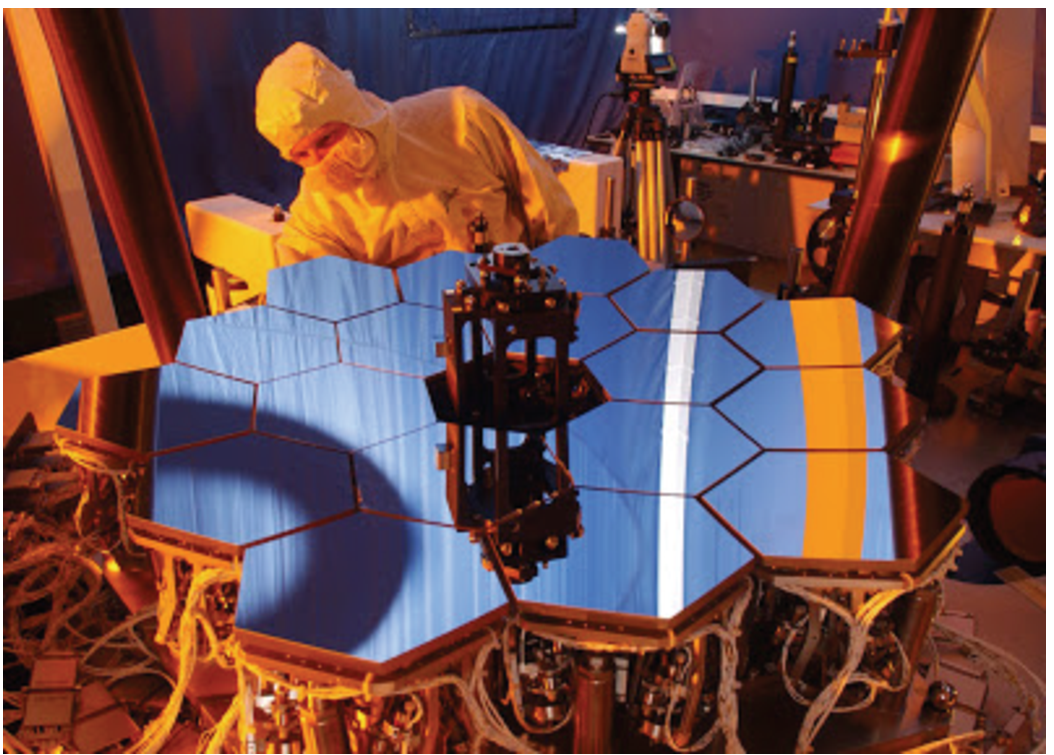}\,\includegraphics[width=4in]{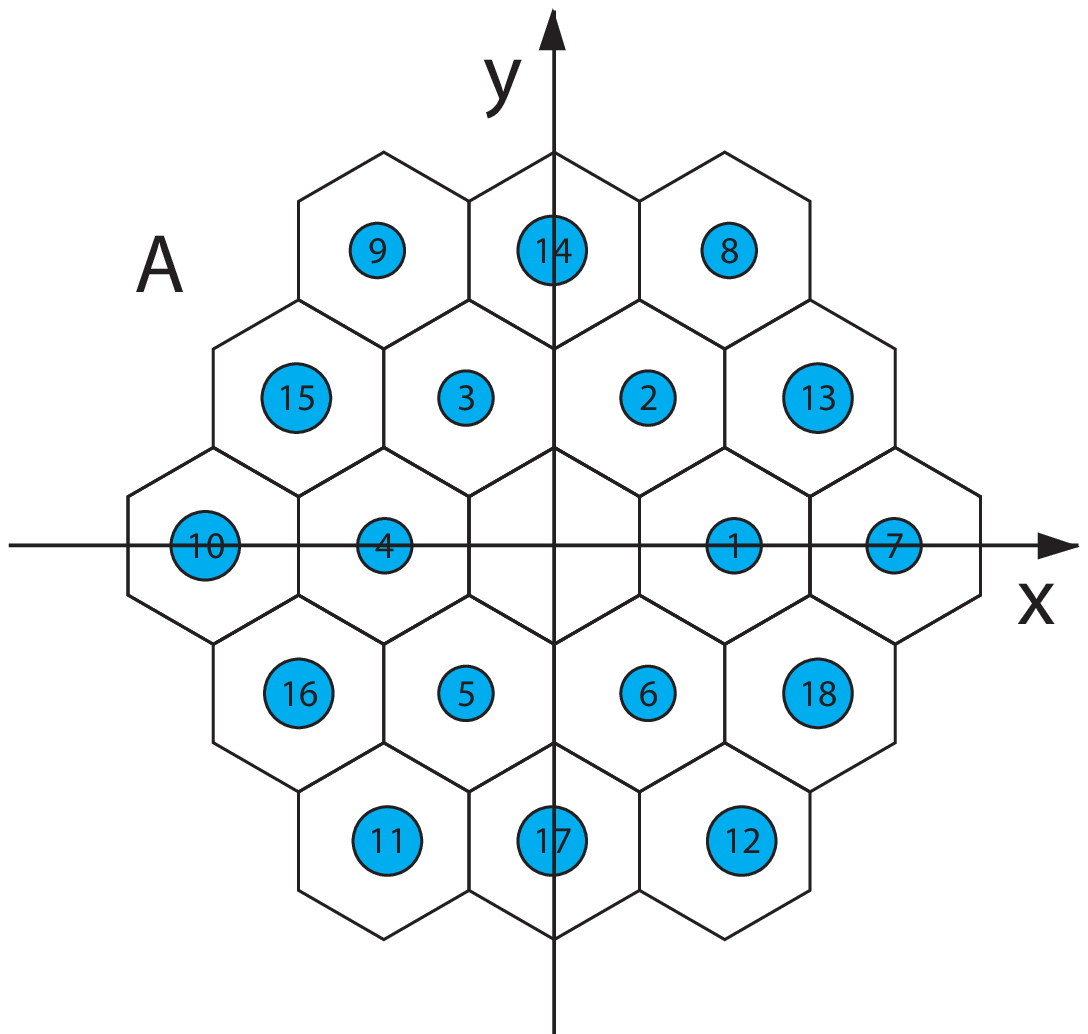}\\
  \caption{Left picture: disposition of the 18 hexagonal facets designed for the primary mirror of the James Webb space telescope projected by NASA (picture taken from http://www.quantum-rd.com/2010/05/el-james-webb-space-telescope-jwst.html). Right picture: disposition of the 18 hexagons that conform the set $A$ with $N=18$ and $p=6$.}
  \end{center}
\end{figure}

Denote by $\lbrace(x_k,y_k)\rbrace_{k=1,2,...,N}$ the cartesian coordinates of the centers of every one of the $N$ polygons $M_k$ of $A$. Then, from the previous section, we know that each one of the $N$ families of functions
$$
[K_n^m]^{(k)}(x,y), \quad k=1,2,\ldots,N,\quad n\in\Ns\cup\{0\}, \hskip 5mm m=-n,-n+2,...,n,
$$
defined by
$$
[K_n^m]^{(k)}(x,y):=K_n^m(x-x_k,y-y_k),
$$
with $K_n^m(x,y)$ given in \eqref{kpolycar} (or \eqref{kpolypol} in polar coordinates), is an orthonormal system in $L^2(M_k)$ with respect to the measure $d\mu_J$ given in \eqref{medida} with $(x,y)$ replaced by $(x-x_k,y-y_k)$.

On the other hand, any function $F$ in the set $L^2(A)$ of complex square integrable functions defined in $A$ may be decomposed in the sum of $N$ functions $F_k\in L^2(M_k)$, each one supported in each one of the polygons $M_k$:
\begin{equation}\label{expansion}
F(x,y)=F(x,y)\sum_{k=1}^{N}\chi_{M_k}(x,y)=\sum_{k=1}^{N}F_k(x,y),
\end{equation}
where $\chi_{M_k}(x,y)$ is the characteristic function of the polygon $M_k$. From the previous section we know that, for every $k=1,2,3,...,N$,
\begin{equation}\label{expansions}
F_k=\sum_{n=0}^\infty\sum_{m=-n}^n [c_n^m]^{(k)}[K_n^m]^{(k)},
\end{equation}
a.e. in $M_k$, with
\begin{equation}\label{coefexpansions}
[c_n^m]^{(k)}={1\over\pi}\int\int_{M_k}K_n^m(x-x_k,y-y_k) F_k(x,y){dxdy\over R_\alpha^2(\arctan((y-y_k)/(x-x_k))}.
\end{equation}
Then we find that
\begin{equation}\label{expansionfunction}
F(x,y)=\sum_{k=1}^{N}\sum_{n=0}^\infty\sum_{m=-n}^n [c_n^m]^{(k)}[K_n^m]^{(k)}(x,y)
\end{equation}
a.e. in $A$. In other words, any complex square integrable function in $A$ may be written, a.e., as a linear combination of elements of the orthonormal system $\lbrace[K_n^m]^{(k)}(x,y)\chi_{M_k}(x,y)\rbrace$ with $k=1,\dots, N$, $n=0,\dots,\infty$ and $m=-n,-n+2,\dots,n$.

In the particular case considered in Figure 6 of an ensemble of $N=18$ hexagons ($p=6$) we have that $\alpha=\pi/6$ and the centers $(x_k,y_k)$ of each one of the 18 hexagons $M_k$ is located at the points
$$
(x_k,y_k)=\sqrt{3}R_0\left(\cos\left({(k-1)\pi\over 3}\right),\sin\left({(k-1)\pi\over 3}\right)\right),\quad 1\leq k\leq 6,
$$
$$
(x_{2k+1},y_{2k+1})=-2\sqrt{3}R_0\left(\cos\left({k\pi\over 3}\right),\sin\left({k\pi\over 3}\right)\right),\quad 3\leq k\leq 8,
$$
$$
(x_{2k},y_{2k})=3R_0\left(-\cos\left({\pi\over 6}-{k\pi\over 3}\right),\sin\left({\pi\over 6}-{k\pi\over 3}\right)\right),\quad 4\leq k\leq 9.
$$

In the appendix we give details about a second example of disposition of hexagonal facets for the primary phase of biger SMT telescope considered in the simulations that we show in the next section.

\section{Atmospheric turbulence simulation and results}
In this section we test the expansion \eqref{expansionfunction} for a realistic simulation of a wavefront distorted by an atmospheric turbulence at the pupil plane of a very large ground based telescope.  It is well-known that the random distortions associated to an atmospheric turbulence follow the Kolmogoroff statistical model \cite{roddier}. Roddier \cite{roddier} proposed an algorithm to generate  realizations of these distorted wavefronts
in terms of Zernike coefficients (or amplitudes). Each realization is obtained by generating normally-distributed random numbers with zero mean and a given covariance matrix according to the Kolmogoroff statistics. We consider the covariance matrix, given in \cite{noll}. This matrix, proposed by Noll from energy considerations, is related to $(D/r_0)^{5/3}$, an overall scaling factor that depends on the telescope's diameter $D$ and the Fried's parameter $r_0$, which represents the coherence length of the wavefront, or the maximum diameter of an ideal diffraction-limited telescope having the same optical resolution (given by the seing disk).

For this simulation, we consider the pupil of the {\it Gran Telescopio Canarias (GTC)}, with a primary segmented mirror that has a diameter of 10.4 metres. It is located in La Palma (2.326 metres of altitude) in the Canary Islands of Spain. The primary mirror consists of 36 hexagonal facets (disposed as indicated in the appendix). The segmented primary mirror is adaptive, which means that both the shape, and relative position of each facet can be changed to compensate atmospheric turbulences, and other possible deformations of the telescope structure. For this reason, it is crucial to determinate, as much accurately as possible, the wavefront in these adaptive optical telescopes.

The simulated random wavefront follows the Kolmogoroff turbulence model. Using the covariance matrix given in \cite{noll}, with $D=10.4 m$, and fried's parameter $r_0=1 m$ (this is a typical value for a reasonably good seing), we have that, in polar coordinates, the fourth order approximation to one particular realization of the random atmospheric wavefront at a certain instant of time is the function, in phase radians,
\begin{equation}\label{wavefront}
\bar f(r,\theta)= \sum_{n=1}^4\sum_{m=-n}^n a_n^m \bar Z_n^m(r,\theta),
\end{equation}
where $\bar Z_n^m\equiv \bar Z_{n}^m(r,\theta)$ are given in \eqref{defpolares}, the coordinates $(r,\theta)$ are the global polar coordinates corresponding to the reference system given in Figure 7 (left) and Figure 11: $x=r\cos\theta$, $y=r\sin\theta$, and the coefficients $a_n^m$ are computed using the covariance matrix given in \cite{noll}.\par

We approximate the input wavefront \eqref{wavefront} by the expansion \eqref{expansionfunction} up to 4th order:
\begin{equation}\label{expansionwavefront}
\bar f(r,\theta)\simeq\bar f^{\rm approx}(r,\theta):=\sum_{k=1}^{36}\sum_{n=0}^4\sum_{m=-n}^n [c_n^m]^{(k)}[\bar K_n^m]^{(k)}(r_k,\theta_k),
\end{equation}
where the Fourier coefficients $[c_n^m]^{(k)}$ are given in \eqref{coefexpansions}, and $(r_k,\theta_k)$ are the local polar coordinates given in the hexagon $k$: $x-x_k=r_k\cos \theta_k,\, y-y_k=r_k\sin \theta_k$.

The function \eqref{wavefront} represents an incident wavefront before it crosses the telescope's pupil. The right hand side of \eqref{expansionwavefront} is an approximation to the wavefront immediately after passing through the telescope pupil. We compute the coefficients $[c_n^m]^{(k)}$ in \eqref{expansionwavefront} in two different ways. On the one hand we obtain $[c_n^m]^{(k)}$ just from their definition: computing the integral \eqref{coefexpansions}. On the other hand, we use the least square approximation method by fitting the values of the wavefront \eqref{wavefront} in a discrete data set given by a uniform hexagonal mesh in the mirror containing 25 points in each hexagon, as detailed in Figure 7 (left).

\begin{figure}[h!]
\begin{center}
\includegraphics[width=2.25in]{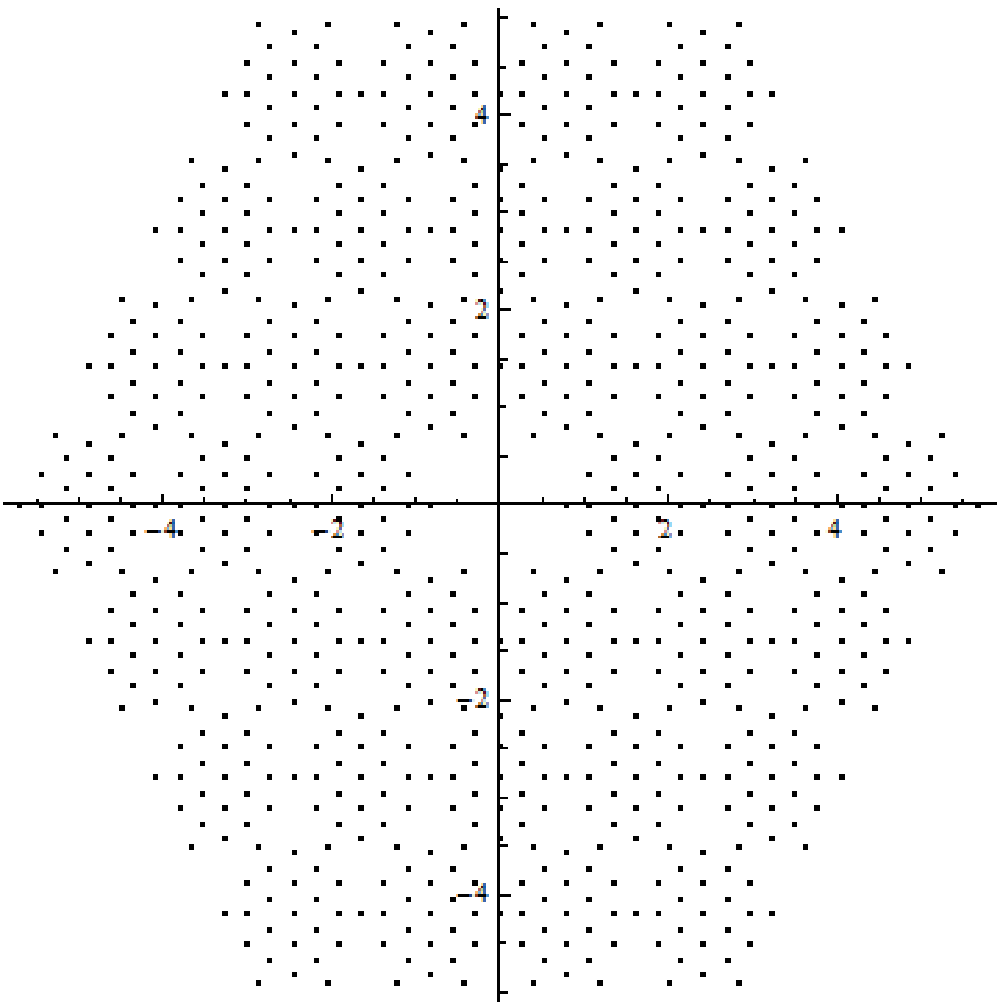}\quad\includegraphics[width=2.25in]{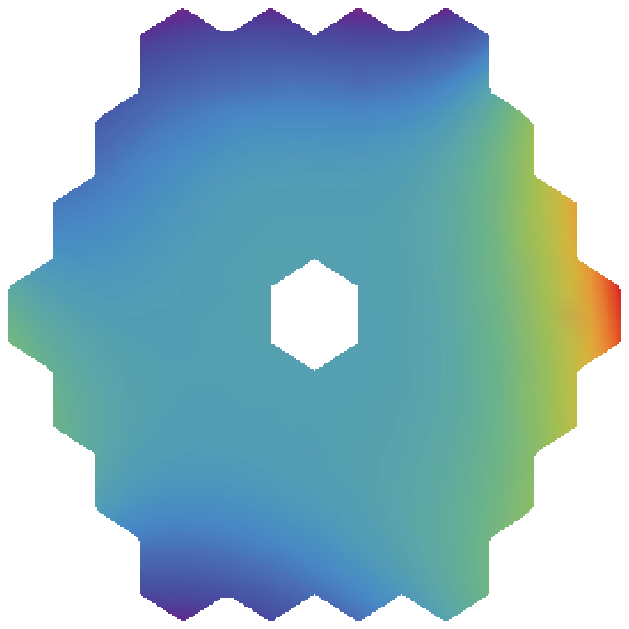}\quad\includegraphics[width=0.4in]{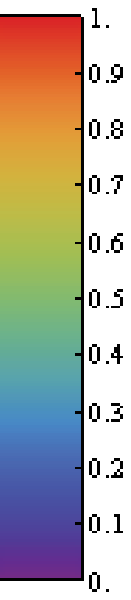}\\
  \caption{Left picture: uniform mesh of the mirror. Right picture: density plot of the wavefront \eqref{wavefront} on the GTC mirror telescope for the particular realization \cite{roddier} $a_n^m=\{$4.25, -9.28, -0.024, -0.11, -0.37, 0.28, -0.75, 2.002, 0.49, 0.35, 0.022, -0.009, 0.34, -0.02$\}$.}
\end{center}
\end{figure}

Therefore, depending on which of the two methods we use to compute the coefficients $[c_n^m]^{(k)}$, we get two different approximations of the function \eqref{wavefront} immediately after passing through the telescope pupil: a ``least square approximation" and ``an integral approximation". The least square approximation allows us to compute approximately the coefficients  $[c_n^m]^{(k)}$ when we do not know the functional expression of the wavefront, but we only know its values at the mesh points. The integral approximation gives us the exact value of the coefficients $[c_n^m]^{(k)}$; however, the calculus of the integral \eqref{coefexpansions} is complicated and requires the knowledge of the functional expression of the wavefront.

The accuracy of both approximations is illustrated in Figures 8 and 9 for the wave front \eqref{wavefront} with the values of the coefficients $a_n$ given in the caption of Figure 7. Figure 8 shows the density plots of the exact function \eqref{wavefront} and the approximation $\bar f^{\rm approx}(r,\theta)$ given in the right hand side of \eqref{expansionwavefront} with coefficients $[c_n^m]^{(k)}$ computed by both methods. We define the relative error of the approximation in the form:
\begin{equation}\label{error}
 f^{\rm re}(r,\theta) := \left| {\bar f(r,\theta)- \bar f^{\rm approx}(r,\theta)\over \bar f(r,\theta)}\right|,
\end{equation}
where $\bar f^{\rm approx}(r_i,\theta_i)$ represents either, the least square, or the integral approximation.
Because of the small values of $\bar{f}^{\rm re}(r,\theta)$, in order to visualize graphically the behavior of this function, we show in Figure 9, its cubic root. We define the relative root mean squared fitting error in the form:
\begin{equation}\label{rmse}
{\rm rrmse}^{\rm approx} :=\displaystyle \left({{1\over M}\sum_{i=1}^M \bar{f}^{\rm re}(r_i,\theta_i)^2}\right)^{1/2},\quad M=\hbox{total number of points of the mesh}.
\end{equation}
In Figure 10 we show ${\rm rrmse}^{\rm approx}$ for some simulations using both methods.

\begin{figure}[h!]
\begin{center}
\includegraphics[width=1.9in]{wavefront_real.eps}
\,\includegraphics[width=1.9in]{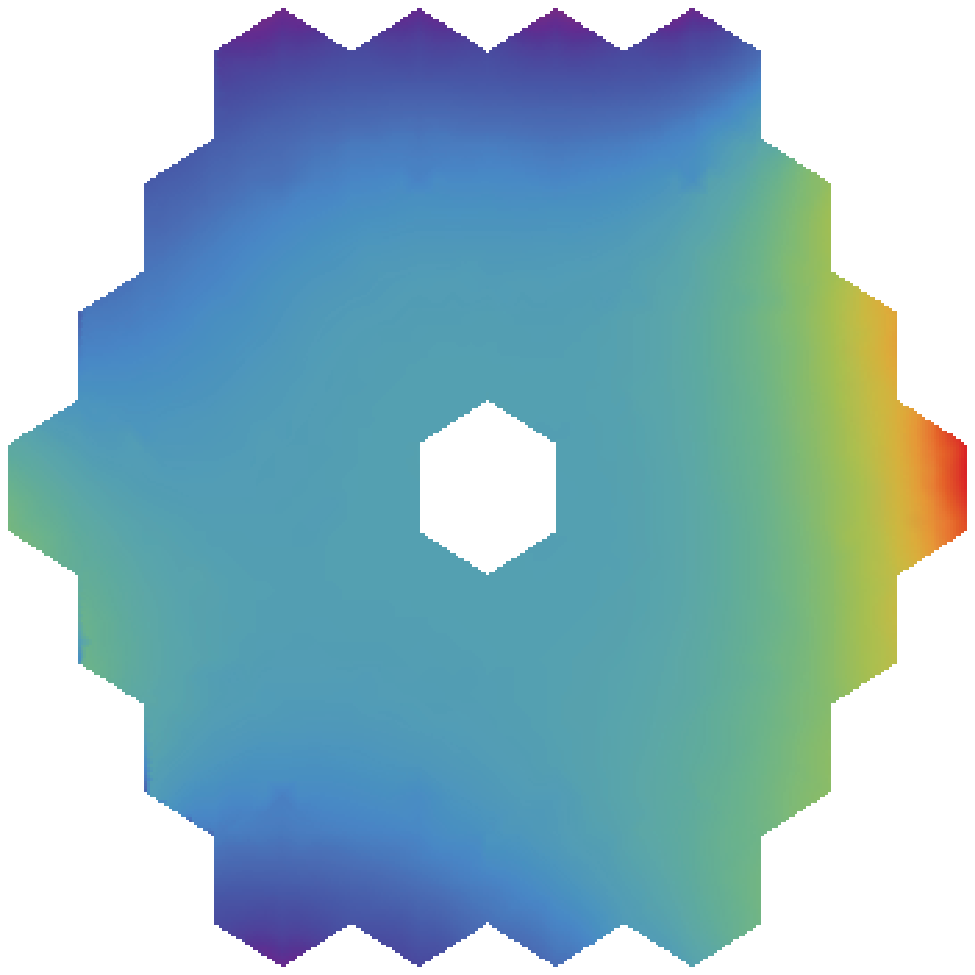}\,\includegraphics[width=1.9in]{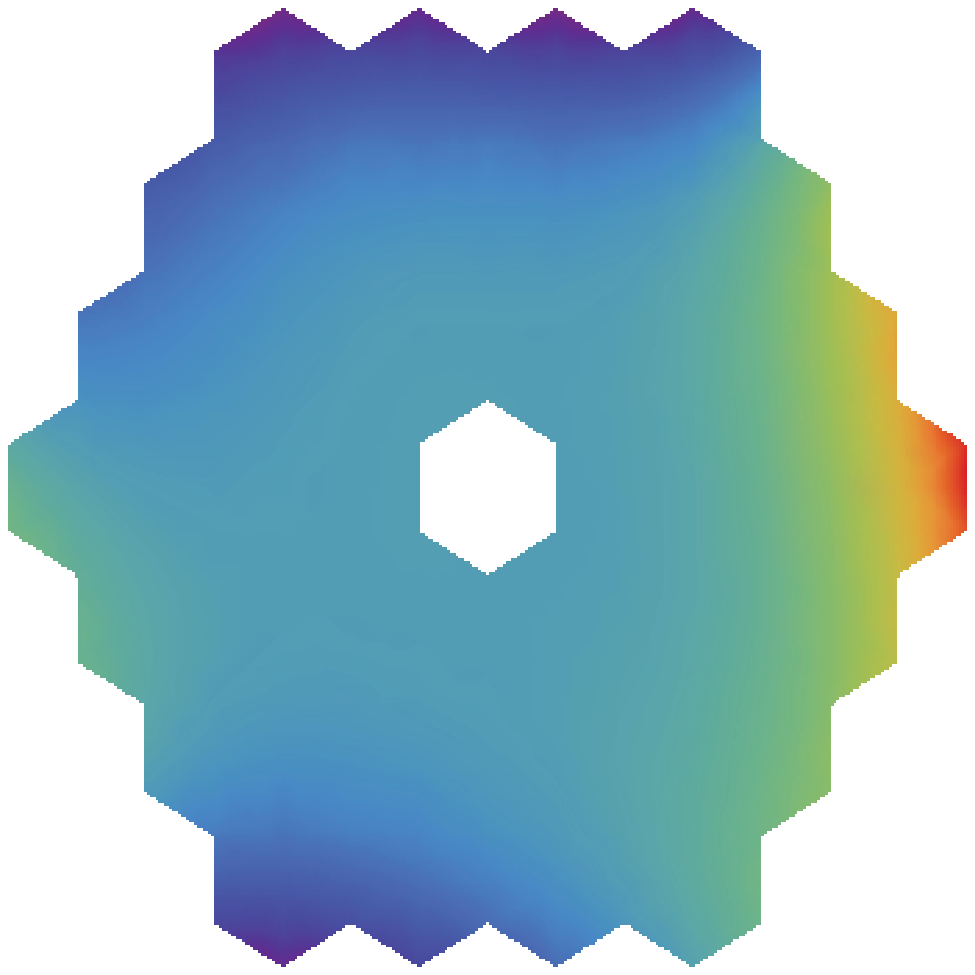}\,\includegraphics[width=0.4in]{colorbar.eps}\\
  \caption{Left picture: function \eqref{wavefront} for the particular coefficients given in the caption of Figure 7. Middle picture: the least square approximation. Right picture: the integral approximation.}
\end{center}
\end{figure}

\begin{figure}[h!]
\begin{center}
\includegraphics[width=1.9in]{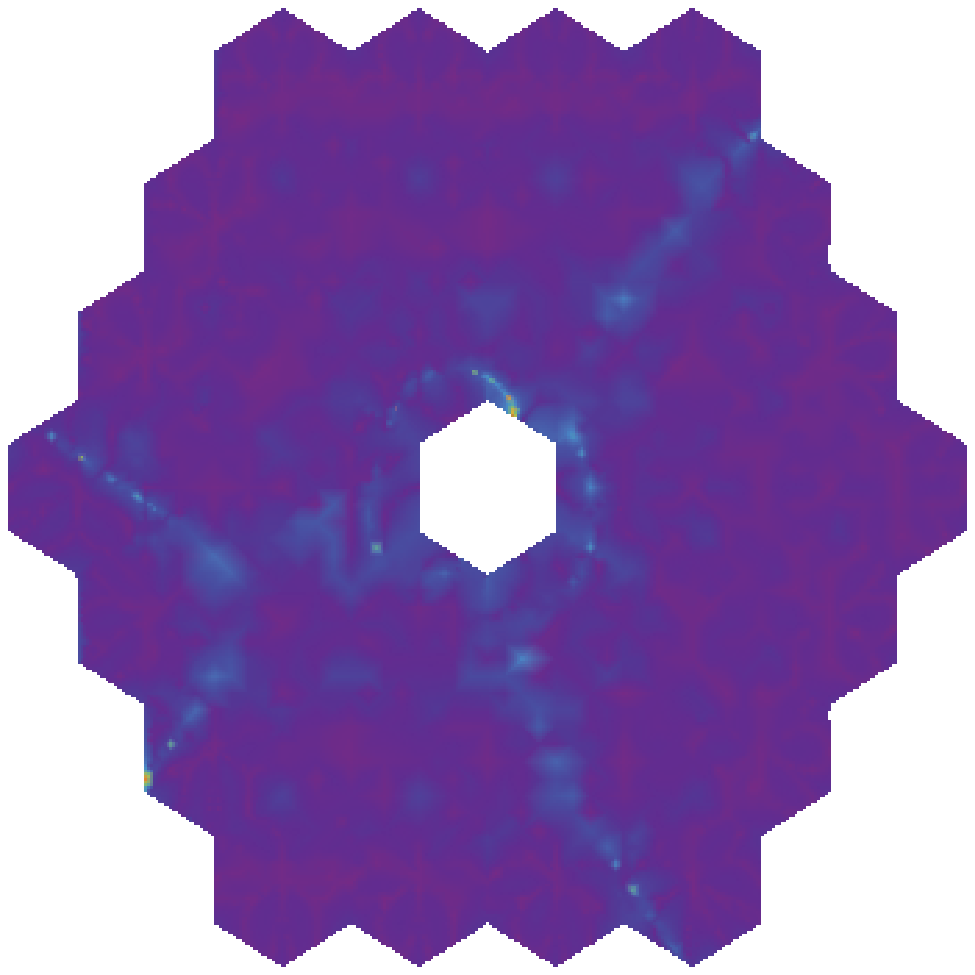}\quad\includegraphics[width=1.9in]{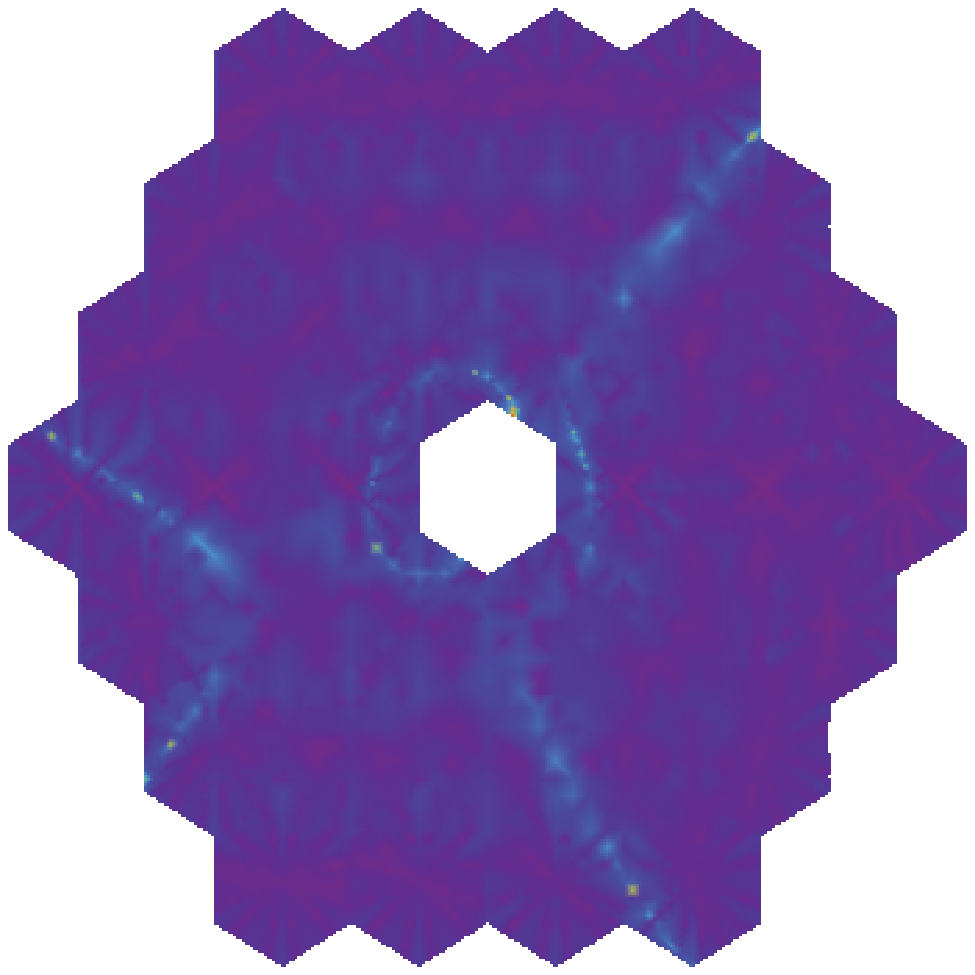}\quad\includegraphics[width=0.4in]{colorbar.eps}\\
  \caption{Cubic root of the relative error function $\bar{f}^{\rm re}(r,\theta)$ given in \eqref{rmse}, for the least square approximation (left), and the integral approximation (right).}
\end{center}
\end{figure}

\begin{figure}[h!]
\begin{center}
\includegraphics[width=2.5in]{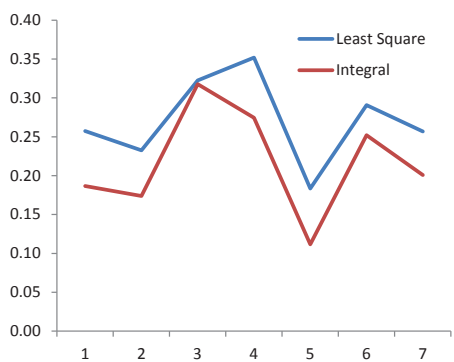}\\
  \caption{Relative root mean squared fitting errors \eqref{rmse}, obtained using the least square and integral approximation for 7 different realizations of \eqref{wavefront} (7 different sets of coefficients $a_n^m$).}
\end{center}
\end{figure}

These results suggest that the accuracy of the estimations is reasonable. The two methods used to compute the coefficients $[c_n^m]^{(k)}$, the least square and the integral method, seem to be equivalent in practice.

\section{Final Remarks}

In this paper we present a rigorous formulation for the general method introduced in \cite{nuestro} to obtain complete orthonormal Zernike-like systems in any set $M$ of the plane piece-wise diffeomorphic to the unit disk. It consists of applying a diffeomorphism to the Zernike system on the unit disk to transform the circle basis functions into the new system on the desired geometry. We then particularize the method to polygons and ensembles of polygonal facets, that are especially important in the wave theory of aberrations and optical image formation in segmented mirror telescopes. In Section 5 we perform a numerical experiment that shows the efficiency of the Zernike-like system obtained in Section 4 for the SMTs.

We believe that the mapping method proposed here, implemented as a change of variables, overcomes most of the difficulties and drawbacks of previous methods, and provides a common framework, especially well-suited for a unified theory of aberrations. On the other hand, the precise physical meaning of the new zernike-like basis $K_n^m(x,y)$ deserves a further investigation.

\section{Acknowledgments}

This research was supported by the Spanish Ministry of Econom\'{\i}a y Competitividad and the
European Union, grant FIS2011-22496, and by the Government of Arag\'{o}n, research group E99. The financial support of the State University of Navarra is also acknowledged.

\section{Appendix}

A second important example frequently used in the primary phase of SMTs telescopes consist of 36 hexagons $M_k$, $k=1,2,...,36$ disposed as in Figure 11.

\begin{figure}[h!]
\begin{center}
\includegraphics[width=5in]{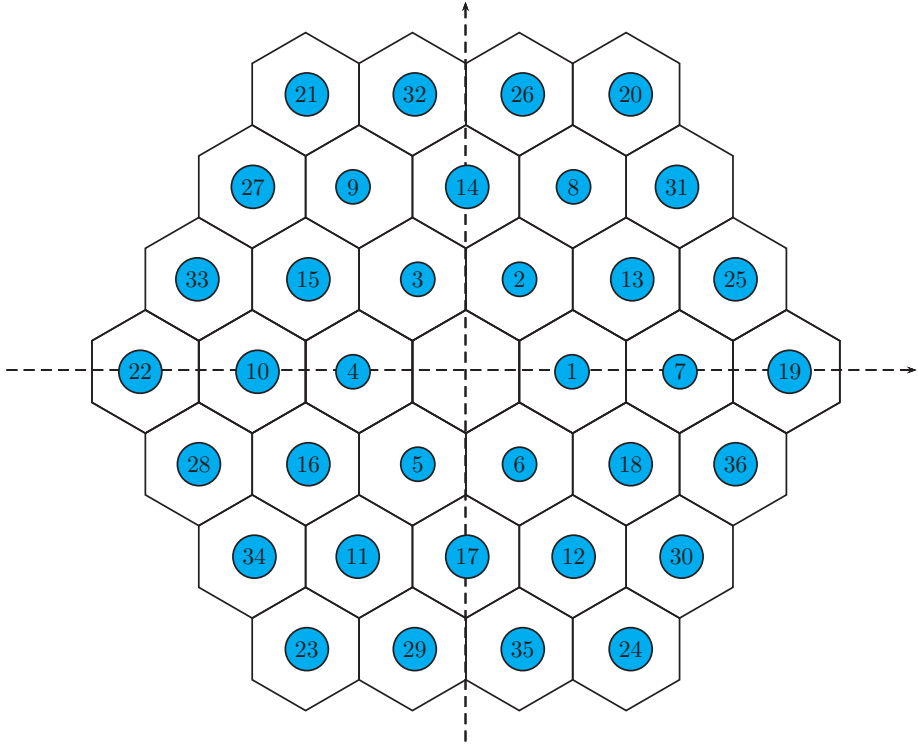}\\
  \caption{Disposition of the 36 hexagons that conform the set $A$ with $N=36$ and $p=6$.}
\end{center}
\end{figure}

The centers $(x_k,y_k)$ of each one of the 36 hexagons $M_k$ is located at the points

$$
(x_k,y_k)=\sqrt{3}R_0\left(\cos\left({(k-1)\pi\over 3}\right),\sin\left({(k-1)\pi\over 3}\right)\right),\quad 1\leq k\leq 6,
$$
$$
(x_{2k-7},y_{2k-7})=2\sqrt{3}R_0\left(\sin\left({\pi\over 6}+{k \pi\over 3}\right),-\cos\left({\pi\over 6}+{k\pi\over 3}\right)\right),\quad 7\leq k\leq 12,
$$
$$
(x_{2k-18},y_{2k-18})=3R_0\left(\cos\left({\pi\over 6}-{k\pi\over 3}\right),-\sin\left({\pi\over 6}-{k\pi\over 3}\right)\right),\quad 13\leq k\leq 18,
$$
$$
(x_{3k-38},y_{3k-38})=3\sqrt{3}R_0\left(\cos\left({(k-19)\pi\over 3}\right),\sin\left({(k-19)\pi\over 3}\right)\right),\quad 19\leq k\leq 24,
$$
$$
(x_{3k-55},y_{3k-55})=\sqrt{21}R_0\left(\cos\left({(k-25)\pi\over 3}+\arctan\left({\sqrt{3}\over 5}\right)\right),
\sin\left({(k-25)\pi\over 3}+\arctan\left({\sqrt{3}\over 5}\right)\right)\right),
$$
$$
\quad 25\leq k\leq 30,
$$
$$
(x_{3k-72},y_{3k-72})=\sqrt{21}R_0\left(\cos\left({(k-31)\pi\over 3}+\arctan\left({\sqrt{3}\over 5}\right)\right),
\sin\left({(k-31)\pi\over 3}+\arctan\left({\sqrt{3}\over 5}\right)\right)\right),
$$
$$
\quad 31\leq k\leq 36.
$$

\end{document}